\documentclass[11pt]{article}

\usepackage{amsfonts,amssymb,chet,dsfont,graphicx,mathrsfs,pgfplots,tikz}

\usetikzlibrary{external}
\title{Polarization Formalism for ALP-induced X-ray Emission from Magnetars}

\author{Jean-Fran\c{c}ois Fortin$^{\ast,}$\email{jean-francois.fortin@phy.ulaval.ca} and Kuver Sinha$^{\dagger,}$\email{kuver.sinha@ou.edu}}

\affiliation{
$^\ast$D\'epartement de Physique, de G\'enie Physique et d'Optique,\\Universit\'e Laval, Qu\'ebec, QC G1V 0A6, Canada\\
$^\dagger$Department of Physics and Astronomy, University of Oklahoma, Norman, OK 73019, USA
}

\abstract{%
Missions like NASA's Imaging X-ray Polarimetry Explorer (IXPE) are poised to provide an unprecedented view of the Universe in polarized X-rays.  Polarization probes physical anisotropies, a fact exploited by particle physicists to look for the anisotropic $a\boldsymbol{E}\cdot\boldsymbol{B}$ operator in the axion-like-particle (ALP) Lagrangian.  Such studies have typically focused on polarization in the radio and microwaves, through local or cosmic birefringence effects.  To such polarization studies we add X-rays emanating from magnetars---a  class of neutron stars with near-critical strength magnetic fields---that are important targets for IXPE.  ALPs produced in the neutron star core convert to X-rays in the magnetosphere; such X-rays are polarized along the direction parallel to the dipolar magnetic field at the point of conversion.  We develop the full theoretical formalism for ALP-induced polarization in the presence of dipolar magnetic fields.  For uncorrelated photon and ALP production mechanisms, we completely disentangle the ALP contributions to the Stokes parameters in terms of the ALP intensity, the ALP-to-photon conversion probability, and the ALP-induced birefringence.  In the proper limit, our results demonstrate that the inclusion of ALPs suppresses the observed degree of circular polarization compared to its pure astrophysical value.  Our results can also be used to impose limits on ALP couplings with IXPE polarization data from magnetars 4U 0142+61 and 1RXS J170849.0-400910, the subject of upcoming work.
}

\date{March 2023} 

\begin{document}

\maketitle

\toc


\section{Introduction}\label{SecIntro}

The investigation of polarized emission in astrophysical settings has become an increasingly important arena for particle physicists.  The reason is the following: polarization probes physical anisotropies, and a particularly important anisotropy occurs in the Lagrangian for axion-like-particles (ALPs) \cite{Weinberg:1977ma,Peccei:1977hh,Wilczek:1977pj,Preskill:1982cy,Dine:1982ah,Kim:1979if}:
\eqn{\mathcal{L}\supset-\frac{g}{4}aF_{\mu\nu}\tilde{F}^{\mu\nu}+g_{aN}(\partial_{\mu}a)\bar{N}\gamma^{\mu}\gamma_5N.}[EqnL]
Here, $a$ denotes the ALP and the couplings $g\equiv g_{a\gamma}$ and $g_{aN}$ denote the ALP-photon and ALP-nucleon couplings, respectively.  The ALP coupling to the photon through the operator $a\boldsymbol{E}\cdot \boldsymbol{B}$ leads to two well-studied polarization effects: $(i)$ in the presence of a background magnetic field, there is the possibility of ALP-photon conversion \cite{Raffelt:1987im,Marsh:2015xka,Graham:2015ouw,Fortin:2018ehg,Fortin:2021cog}.  Since the ALP only couples to the parallel component of the electric field, such conversion necessarily gives polarized photons; and $(ii)$ in the presence of an ALP background, photons propagate through what is essentially a birefringent medium.  This, again, results in various effects: for a time-dependent ALP background, one would have a rotation of the plane of polarized light.  Such effects are being searched for in the cosmic microwave background (the ``cosmic birefringence effect") \cite{Carroll:1989vb,Harari:1992ea,Lue:1998mq} as well as in localized settings around black holes \cite{Chen:2021lvo,Chen:2022oad} (this relies on ALP clouds existing due to super-radiance).

Our interest in this paper is in the first category of effects: polarization resulting from ALP-photon conversions.  The conversions relevant for us are localized around neutron stars.  This should be contrasted with the widely studied effect of conversions of ALPs in long-distance galactic and inter-galactic magnetic fields; while such long distance conversions are interesting, there are significant challenges in modeling the magnetic field and the diverse environments the ALP must traverse \cite{Conlon:2018iwn,Buehler:2020qsn, Calore:2020tjw,Payez:2014xsa,Xiao:2020pra,Kartavtsev:2016doq,Day:2015xea,Dobrynina:2014qba}.  Conversion in the vicinity of compact objects like neutron stars or white dwarfs occurs in a much more controlled environment, holding out the promise of precision studies.

Within the class of such local ALP-photon conversions, too, there is a further division into two categories of models and signatures: $(a)$ cold ALPs and radio signals \cite{Pshirkov:2007st,Hook:2018iia}: in this class of studies, cold ambient ALPs convert to radio photons near neutron stars.  A complication of such scenarios is that one generally has to be careful about the modeling of the plasma near the neutron star; and $(b)$ hot ALPs and X-ray signals \cite{Fortin:2021sst,Harris:2020qim,Lloyd:2020vzs}: in this class of models, relativistic ALPs are produced from the core of the neutron star, and convert to X-ray photons as they travel outwards, typically around a distance of $\mathcal{O}(1000\,r_0)$ ($r_0$ is the neutron star radius).  Since the ALPs in this class of studies are relativistic, the conversion process is independent of the details of the plasma.

The purpose of this paper is to study the effects of ALPs on the polarization of X-rays emanating from neutron stars.  The general mechanism is as follows: ALPs are produced in the core by nucleon bremsstrahlung \cite{Iwamoto:1984ir,Nakagawa:1987pga,Nakagawa:1988rhp,Iwamoto:1992jp,Raffelt:1996wa,Umeda:1997da,Maruyama:2017xzl,Paul:2018msp}; they travel outside and convert to X-ray photons in the strong dipolar magnetic field in the magnetosphere.  These ALP-induced photons are polarized along the direction parallel to the magnetic field at the point of conversion.  The overall polarization pattern of the emission from neutron stars then combines the polarization of photons produced by (non-ALP) astrophysical processes, superposed with the polarization of exclusively ALP-induced photons.  The formal study of the ALP-induced polarization involves solving for the spatial evolution of the Stokes parameters of the ALP-photon system in a dipolar magnetic field.  This study was initiated by the present authors in \cite{Fortin:2018aom}, where expressions for the Stokes parameters $I$ and $Q$ were derived.  In the current study, we complete the formalism by deriving the full set of Stokes parameters, including a treatment of circular polarization.

The scope of our current work is to develop the formal infrastructure based on which the physics of ALPs can be connected to data from polarization observations of neutron stars.  Before proceeding, however, we make a few comments about the observational status of the field, as well as the challenges associated with claiming a signal in the case of an anomaly in the future (in contrast to setting upper limits on ALP couplings).  X-ray polarimetry as an observational discipline has a long history \cite{fabiani} and is poised to enter an era of explosive growth and data \cite{Krawczynski:2019ofl,polstar,ixpe}.  The recent launch of NASA's Imaging X-ray Polarimetry Explorer (IXPE) mission and the subsequent polarization data from magnetars 4U 0142+61 \cite{Taverna:2022jgl} and 1RXS J170849.0-400910 \cite{Zane:2023khc} give us an unprecedented opportunity of bringing questions of fundamental physics to this field.  Magnetars, with their extreme magnetic fields around the quantum critical value $B_c=m_e^2/\sqrt{4\pi\alpha}\approx4.414\times10^{13}\,\text{G}$ \cite{Turolla:2015mwa,Beloborodov:2016mmx,Kaspi:2017fwg}, are particularly important players, since ALP-induced effects grow with the strength of the magnetic field.  The signatures of the ALP-induced polarization also grow parametrically with the coupling $g_{aN}$ of ALPs to nucleons (which controls the production) and the coupling $g$ to photons (which controls the conversion).

The discrimination of ALP-induced polarization from astrophysical ``background" polarization is a very difficult question, since it depends on the astrophysical modeling of the thermal and non-thermal emission \cite{Lai:2006af,Chelouche:2008ta,Jimenez:2011pg,Perna:2012wn,Wadiasingh:2017rcq}.  Non-linear QED effects \cite{gnedin2} and the anisotropic opacities of the surface plasma \cite{Lai:2003nd,Taverna:2015vpa} must be properly taken into account.  The challenges of modeling polarization in the hard X-ray regime (where the emission possibly comes from resonant inverse Compton scattering of the thermal emission) are also significant \cite{Beloborodov:2012ug,Wadiasingh:2017rcq}.  A more pragmatic strategy is to impose reliable upper limits on ALP couplings, by requiring that ALP-induced effects broadly do not supersede the concordance between data and astrophysical modeling.  In a companion paper \cite{upcoming}, this is the strategy that we will pursue to extract the first limits on ALP couplings from IXPE data.

For the convenience of readers who want to directly use our results to compare with data, we summarize the most important expressions here.  In the presence of ALP-photon conversions, the polarization invariants (total intensity $I$, total polarization degree $p$, degree of circular polarization $p_c$, and degree of linear polarization $p_l$) take the following form:
\eqn{
\begin{gathered}
I_\text{obs}\approx(1+p_a)I_\text{astro},\qquad\qquad p_\text{obs}^2\approx\frac{p_\text{astro}^2-2p_aq+p_a^2}{(1+p_a)^2},\\
p_{c,\text{obs}}^2\approx\frac{p_{c,\text{astro}}^2}{(1+p_a)^2},\qquad\qquad p_{l,\text{obs}}^2\approx\frac{p_{l,\text{astro}}^2-2p_aq+p_a^2}{(1+p_a)^2},
\end{gathered}
}
where
\eqn{q=\frac{Q_\text{astro}}{I_\text{astro}},\qquad\qquad p_a=\frac{I_a(1)P_{a\to\gamma}}{I_\text{astro}}.}
These equations give the observed polarization invariants (with subscript ``obs") in terms of the astrophysical polarization invariants (with subscript ``astro") and Stokes parameter $Q_\text{astro}$.  The former is obtained directly from data, while the latter is obtained from the preferred astrophysical model, to which we are agnostic.  The ALP contribution is entirely contained in the parameter $p_a$.  To evaluate $p_a$, two quantities are required: $I_a(1)$, which is the intensity of ALPs produced from the neutron star; and $P_{a\to\gamma}$, which is the probability that an ALP converts to a photon in the magnetosphere.  A conservative estimate for $I_a(1)$ is to take it to be bounded by the intensity of neutrinos produced by the neutron star.  The semi-analytic expression for $P_{a\to\gamma}$, which is an excellent approximation to the full solution for couplings $g$ of interest, is given in \eqref{EqnAllInt}.

Our paper is organized as follows.  Section \ref{SecPolGen} reviews the evolution equations for the ALP-photon coupled systems in the weak dispersion limit, based on our previous (partial) work on polarization \cite{Fortin:2018aom}.  The full set of four Stokes parameters is discussed and two auxiliary quantities are introduced to show that all the Stokes parameters are independent of the initial mixture of ALPs and photons when they are produced without correlations.  This observation allows us to completely disentangle the effects of ALPs on the Stokes parameters in terms of the ALP intensity, the ALP-to-photon conversion probability, and the ALP-induced birefringence.  The results are then expressed in terms of the standard polarization invariants, \textit{i.e.}\ the total intensity $I_\text{obs}$, the degree of total polarization $p_\text{obs}^2$, the degree of circular polarization $p_{c,\text{obs}}^2$, and the degree of linear polarization $p_{l,\text{obs}}^2$.  Section \ref{SecPolPert} then investigates the perturbative regime, where approximate analytical solutions to the ALP-to-photon conversion probability and ALP-induced birefringence are obtained with the help of standard perturbation theory.  The approximations are compared with the numerical solutions to the evolution equations, showing a good match in the proper limit.  We then use these approximations with small ALP-photon coupling to re-express the polarization invariants more simply.  Finally, Section \ref{SecConc} presents our conclusions.


\section{Polarization: General Theory}\label{SecPolGen}

In this section we summarize the general theory of ALP-photon oscillations and introduce the relevant evolution equations.  For concreteness, we focus on the case of magnetars for which magnetic fields are extreme.  Our analysis is performed in the limits of large space variations of the magnetic field when compared to the particle wavelength and weak dispersion \cite{Raffelt:1987im}, where we can proceed with the general oscillation formalism developed in \cite{Fortin:2018ehg}.


\subsection{Evolution Equations}

The evolution equations for photons and ALPs with energies below the electron mass propagating radially outwards from a stellar object can be expressed as \cite{Raffelt:1987im,Lai:2006af}
\eqn{i\frac{d}{dx}\left(\begin{array}{c}a\\E_\parallel\\E_\perp\end{array}\right)=\left(\begin{array}{ccc}\omega r_0+\Delta_ar_0&\Delta_Mr_0&0\\\Delta_Mr_0&\omega r_0+\Delta_\parallel r_0&0\\0&0&\omega r_0+\Delta_\perp r_0\end{array}\right)\left(\begin{array}{c}a\\E_\parallel\\E_\perp\end{array}\right),}[EqnDiffMat]
in the correct limits,\footnote{The two limits are: $(i)$ the limit where variations of the magnetic field occur on distances much larger than the Compton wavelength of the particles; and $(ii)$ the limit of weak dispersion where the refractive indices are close to unity.  For example, for the magnetar's magnetosphere the weak dispersion limit implies that the magnetic field is much smaller than approximatively $\sqrt{\frac{45\pi}{\alpha}}B_c\approx6\times10^{15}\,\text{G}$, where $B_c$ is the critical QED magnetic field strength (see below).} where
\eqn{\Delta_a=-\frac{m_a^2}{2\omega},\qquad\qquad\Delta_\parallel=\frac{1}{2}q_\parallel\omega\sin^2\theta,\qquad\qquad\Delta_\perp=\frac{1}{2}q_\perp\omega\sin^2\theta,\qquad\qquad\Delta_M=\frac{1}{2}gB\sin\theta.}
Here, the $\Delta_M$ contribution originates from the Lagrangian \eqref{EqnL} and \eqref{EqnDiffMat} is valid as long as the plasma contributions are negligible.  Moreover, the fields $a(x)$, $E_\parallel(x)$ and $E_\perp(x)$ are the ALP, parallel and perpendicular photon electric fields, respectively.  They are functions of the dimensionless distance from the magnetar center 
\eqn{x=r/r_0,}
with $r$ the distance from the center of the magnetar and $r_0$ the magnetar's radius.\footnote{Throughout the paper, the presence of the subscript $0$ indicates that the related quantity is evaluated at the surface, \textit{e.g.}\ $\Delta_{M0}=\left.\Delta_M\right|_{x=1}$.}  The particle's energy is denoted by $\omega$, the ALP mass by $m_a$, the ALP-photon coupling constant by $g$, and the angle between the magnetic field and the direction of propagation of the particle by $\theta$.  Finally, the effect of the magnetosphere is encoded in $q_\parallel$ and $q_\perp$ which are dimensionless functions of the magnetic field $B$ given by \cite{Lai:2006af,Raffelt:1987im}
\eqn{
\begin{gathered}
q_\parallel=\frac{7\alpha}{45\pi}b^2\hat{q}_\parallel,\qquad\qquad\hat{q}_\parallel=\frac{1+1.2b}{1+1.33b+0.56b^2},\\
q_\perp=\frac{4\alpha}{45\pi}b^2\hat{q}_\perp,\qquad\qquad\hat{q}_\perp=\frac{1}{1+0.72b^{5/4}+(4/15)b^2},
\end{gathered}
}
with $b=B/B_c$.  Here $B_c=m_e^2/\sqrt{4\pi\alpha}\approx4.414\times10^{13}\,\text{G}$ is the critical QED magnetic field strength expressed in terms of the electron mass $m_e$ and the fine structure constant $\alpha$.

It is important to note that the absence of plasma contributions (which are completely negligible everywhere except in the centimeter-thick plasma around the magnetar \cite{Lai:2006af}) leads to a factorization of the three-state system \eqref{EqnDiffMat} into a two-state system for the ALP and parallel photon fields, and a one-state system for the perpendicular photon field.

The probability conservation property discussed in \cite{Fortin:2018ehg} implies that $\frac{d}{dx}[|a(x)|^2+|E_\parallel(x)|^2]=0$, from which it is possible to express the different states as
\eqn{a(x)=A\cos[\chi(x)]e^{-i\phi_a(x)},\qquad\qquad E_\parallel(x)=iA\sin[\chi(x)]e^{-i\phi_\parallel(x)},\qquad\qquad E_\perp(x)=A_\perp e^{-i\phi_\perp(x)}.}[EqnaEE]
Thus, the ALP and photon fields amplitudes at position $xr_0$ are $A_a=A\cos[\chi(x)]$, $A_\parallel=A\sin[\chi(x)]$ and $A_\perp$, with the intensities at position $xr_0$ given by the respective amplitudes squared, \textit{i.e.}\ $I_a(x)=A^2\cos^2[\chi(x)]$, $I_\parallel(x)=A^2\sin^2[\chi(x)]$ and $I_\perp(x)=A_\perp^2$.  Here $A$ and $A_\perp$ are constants that can be chosen real and positive while $\chi(x)$, $\phi_a(x)$, $\phi_\parallel(x)$ and $\phi_\perp(x)$ are real functions.  At the level of intensities, the probability conservation property corresponds to $I_a(x)+I_\parallel(x)=A^2$ a constant, with $I_\perp(x)=A_\perp^2$ also a constant.

Using \eqref{EqnaEE} in \eqref{EqnDiffMat}, the evolution equations become
\eqn{
\begin{gathered}
\frac{d\chi(x)}{dx}=-\Delta_Mr_0\cos[\Delta\phi(x)],\\
\frac{d\Delta\phi(x)}{dx}=(\Delta_a-\Delta_\parallel)r_0+2\Delta_Mr_0\cot[2\chi(x)]\sin[\Delta\phi(x)],\\
\frac{d\delta\phi(x)}{dx}=(\Delta_\perp-\Delta_\parallel)r_0+\Delta_Mr_0\{\cot[2\chi(x)]+\csc[2\chi(x)]\}\sin[\Delta\phi(x)],\\
\frac{d\phi_\perp(x)}{dx}=(\omega+\Delta_\perp)r_0,
\end{gathered}
}[EqnEE]
where 
\eqn{\Delta\phi(x)=\phi_a(x)-\phi_\parallel(x),\qquad\qquad\delta\phi(x)=\phi_\perp(x)-\phi_\parallel(x),}[EqnDeltaPhi]
are the phase difference between the ALP field and the parallel photon field and the phase difference between the two photon field polarizations, respectively.  We note that the differential equation for $\phi_\perp(x)$ decouples from the evolution equations \eqref{EqnEE} while the differential equation for $\delta\phi(x)$ is completely determined once the solutions to the coupled differential equations for $\chi(x)$ and $\Delta\phi(x)$ are known.


\subsection{Stokes Parameters}

In magnetars, the production of X-ray photons (both soft and hard) and the production of ALPs do not have the same origin and are uncorrelated.  ALP production mainly comes from nucleon-nucleon bremsstrahlung of ALPs in the core of the magnetars \cite{Iwamoto:1984ir,Iwamoto:1992jp,Nakagawa:1987pga,Nakagawa:1988rhp,Raffelt:1996wa,Umeda:1997da,Paul:2018msp,Maruyama:2017xzl}, and the ALP-induced X-ray photons are polarized along the parallel direction.  On the other hand, the photons coming from astrophysical (non-ALP) processes have a different pattern of polarization, which depends on the modeling.

We make a few comments about the polarization of astrophysical photons, briefly summarizing the literature.  In general, the extraordinary or perpendicular X-mode opacity is suppressed (enhanced) compared to the ordinary or parallel O-mode opacity in the soft (hard) emission \cite{Lai:2003nd}, implying emission mostly in the X-mode (O-mode) in the soft (hard) regime.  However, there are further subtleties depending on the plasma.  For example, the presence of a centimeter-thick plasma can lead to a vacuum resonance occurring from its interplay with vacuum polarization \cite{Lai:2006af}.  This vacuum resonance can convert X-mode and O-mode into each other, leading to a change in the dominant polarization mode.  On the one hand, the soft emission, which is thought to originate from thermal photons produced by the magnetar's surface, is mostly in the extraordinary polarization mode.  On the other hand, for magnetars with $B_0\gtrsim7\times10^{14}\,\text{G}$, the non-thermal emission is also thought to be dominated by the extraordinary polarization mode due to the vacuum resonance in the inhomogeneous plasma \cite{Lai:2006af}.  For such magnetars, then, the entire astrophysical emission spectrum, both soft and hard, could be predominantly polarized in the  extraordinary mode.  We note that hard X-ray production is still under active study and could come from several mechanisms.  One such mechanism might be resonant inverse Compton scattering of soft X-ray photons by ultrarelativistic charged particles in the magnetosphere \cite{Wadiasingh:2017rcq}.  Another mechanism considers relativistic particle injection in the magnetosphere \cite{Beloborodov:2012ug}.  In both cases, it was found that the resulting hard X-ray spectrum is strongly polarized in the X-mode.

Since processes responsible for the creation of ALPs and astrophysical photons (in whichever mode they are polarized) are unrelated, the phase difference between the ALP field and parallel photon field is arbitrary, and it is natural to average over the phase difference at the surface $\Delta\phi_0$.  Therefore, the Stokes parameters are
\eqna{
I(\chi_0,x)&=\int_0^{2\pi}\frac{d\Delta\phi_0}{2\pi}\,[|E_\perp(x)|^2+|E_\parallel(\chi_0,x)|^2]\\
&=A_\perp^2+A^2\int_0^{2\pi}\frac{d\Delta\phi_0}{2\pi}\,\sin^2\left[\left.\chi(x)\right|_{\chi(1)=\chi_0,\Delta\phi(1)=\Delta\phi_0}\right],\\
Q(\chi_0,x)&=\int_0^{2\pi}\frac{d\Delta\phi_0}{2\pi}\,[|E_\perp(x)|^2-|E_\parallel(\chi_0,x)|^2],\\
&=A_\perp^2-A^2\int_0^{2\pi}\frac{d\Delta\phi_0}{2\pi}\,\sin^2\left[\left.\chi(x)\right|_{\chi(1)=\chi_0,\Delta\phi(1)=\Delta\phi_0}\right],\\
U(\chi_0,\delta\phi_0,x)&=\int_0^{2\pi}\frac{d\Delta\phi_0}{2\pi}\,[E_\perp(x)E_\parallel(\chi_0,x)^*+E_\perp(x)^*E_\parallel(\chi_0,x)],\\
&=-2A_\perp A\int_0^{2\pi}\frac{d\Delta\phi_0}{2\pi}\,\sin\left[\left.\chi(x)\right|_{\chi(1)=\chi_0,\Delta\phi(1)=\Delta\phi_0}\right]\sin\left[\delta\phi_0+\left.\delta\phi(x)\right|_{\chi(1)=\chi_0,\Delta\phi(1)=\Delta\phi_0}\right],\\
V(\chi_0,\delta\phi_0,x)&=\int_0^{2\pi}\frac{d\Delta\phi_0}{2\pi}\,i[E_\perp(x)E_\parallel(\chi_0,x)^*-E_\perp(x)^*E_\parallel(\chi_0,x)],\\
&=2A_\perp A\int_0^{2\pi}\frac{d\Delta\phi_0}{2\pi}\,\sin\left[\left.\chi(x)\right|_{\chi(1)=\chi_0,\Delta\phi(1)=\Delta\phi_0}\right]\cos\left[\delta\phi_0+\left.\delta\phi(x)\right|_{\chi(1)=\chi_0,\Delta\phi(1)=\Delta\phi_0}\right],
}[EqnStokes]
where it is understood that $\delta\phi(1)=0$ since the dependence on the phase difference at the surface $\delta\phi_0$ has been extracted explicitly.  We stress again that a subscript $0$ indicates that the corresponding quantity is evaluated at the magnetar's surface.  The surface-subtracted quantities thus become
\eqn{
\begin{gathered}
I(\chi_0,x)-I(\chi_0,1)=A^2\cos(2\chi_0)\int_0^{2\pi}\frac{d\Delta\phi_0}{2\pi}\,P(\chi_0,\Delta\phi_0,x),\\
Q(\chi_0,x)-Q(\chi_0,1)=-[I(\chi_0,x)-I(\chi_0,1)],\\
U(\chi_0,\delta\phi_0,x)-U(\chi_0,\delta\phi_0,1)=2A_\perp A\sin(\chi_0)\Im\left\{e^{-i\delta\phi_0}\int_0^{2\pi}\frac{d\Delta\phi_0}{2\pi}\,S(\chi_0,\Delta\phi_0,x)\right\},\\
V(\chi_0,\delta\phi_0,x)-V(\chi_0,\delta\phi_0,1)=2A_\perp A\sin(\chi_0)\Re\left\{e^{-i\delta\phi_0}\int_0^{2\pi}\frac{d\Delta\phi_0}{2\pi}\,S(\chi_0,\Delta\phi_0,x)\right\},
\end{gathered}
}[EqnStokesDiff]
where we introduced
\eqn{
\begin{gathered}
P(\chi_0,\Delta\phi_0,x)=\frac{1}{2}\left\{1-\frac{\cos\left[\left.2\chi(x)\right|_{\chi(1)=\chi_0,\Delta\phi(1)=\Delta\phi_0}\right]}{\cos(2\chi_0)}\right\},\\
S(\chi_0,\Delta\phi_0,x)=\frac{\sin\left[\left.\chi(x)\right|_{\chi(1)=\chi_0,\Delta\phi(1)=\Delta\phi_0}\right]}{\sin(\chi_0)}e^{-i\!\left.\delta\phi(x)\right|_{\chi(1)=\chi_0,\Delta\phi(1)=\Delta\phi_0}}-1.
\end{gathered}
}[EqnPS]
The quantity $P(\chi_0,\Delta\phi_0,x)$, which will relate to the ALP-to-photon conversion probability, was already introduced in \cite{Fortin:2018aom}.  The quantity $S(\chi_0,\Delta\phi_0,x)$, which will lead to ALP-induced birefringence, is new. 


\subsection{Evolution of the Auxiliary Quantities}

In this subsection, we study the evolution of $P(\chi_0,\Delta_0,x)$ and $S(\chi_0,\Delta_0,x)$ to show that their $\Delta\phi_0$-average is independent of $\chi_0$.  We achieve this from the evolution equations \eqref{EqnEE} rewritten as
\eqn{
\begin{gathered}
\frac{d\chi(x)}{dx}=-D(x)\cos[\Delta\phi(x)],\\
\frac{d\Delta\phi(x)}{dx}=C(x)+2D(x)\cot[2\chi(x)]\sin[\Delta\phi(x)],\\
\frac{d\delta\phi(x)}{dx}=E(x)+D(x)\{\cot[2\chi(x)]+\csc[2\chi(x)]\}\sin[\Delta\phi(x)].
\end{gathered}
}
It is easy to verify that the quantities defined in \eqref{EqnPS} satisfy the following differential equations
\eqna{
0&=\frac{d^3}{dx^3}P(\chi_0,\Delta\phi_0,x)-\ln[C(x)D(x)^2]'\frac{d^2}{dx^2}P(\chi_0,\Delta\phi_0,x)\\
&\phantom{=}\qquad+\{C(x)^2+4D(x)^2+\ln[C(x)D(x)]'\ln[D(x)]'-\ln[D(x)]''\}\frac{d}{dx}P(\chi_0,\Delta\phi_0,x)\\
&\phantom{=}\qquad+2D(x)^2\ln[D(x)/C(x)]'[2P(\chi_0,\Delta\phi_0,x)-1],\\
0&=\frac{d^2}{dx^2}S(\chi_0,\Delta\phi_0,x)-\{iC(x)-2iE(x)+\ln[D(x)]'\}\frac{d}{dx}S(\chi_0,\Delta\phi_0,x)\\
&\phantom{=}\qquad+\{D(x)^2+C(x)E(x)-E(x)^2-iE(x)\ln[D(x)/E(x)]'\}[S(\chi_0,\Delta\phi_0,x)+1],
}[EqnPSdiff]
with boundary conditions
\eqn{
\begin{gathered}
P(\chi_0,\Delta\phi_0,x=1)=0,\\
\left.\frac{d}{dx}P(\chi_0,\Delta\phi_0,x)\right|_{x=1}=-D(1)\tan(2\chi_0)\cos(\Delta\phi_0),\\
\left.\frac{d^2}{dx^2}P(\chi_0,\Delta\phi_0,x)\right|_{x=1}=2D(1)^2+C(1)D(1)\tan(2\chi_0)\sin(\Delta\phi_0)-D'(1)\tan(2\chi_0)\cos(\Delta\phi_0),\\
S(\chi_0,\Delta\phi_0,x=1)=0,\\
\left.\frac{d}{dx}S(\chi_0,\Delta\phi_0,x)\right|_{x=1}=-iE(1)-D(1)\cot(\chi_0)e^{i\Delta\phi_0}.
\end{gathered}
}[EqnPSbnd]
Here a prime denotes a derivative with respect to $x$.  Clearly, $P(\chi_0,\Delta\phi_0,x)$ and $S(\chi_0,\Delta\phi_0,x)$ depend on $\chi_0$ and $\Delta\phi_0$ through their boundary conditions \eqref{EqnPSbnd}, but not through their differential equations \eqref{EqnPSdiff}.  Moreover, since the differential equations \eqref{EqnPSdiff} are linear in $P(\chi_0,\Delta\phi_0,x)$ and $S(\chi_0,\Delta\phi_0,x)$ respectively, the averaged quantities
\eqn{\bar{P}(\chi_0,x)=\int_0^{2\pi}\frac{d\Delta\phi_0}{2\pi}\,P(\chi_0,\Delta\phi_0,x),\qquad\qquad\bar{S}(\chi_0,x)=\int_0^{2\pi}\frac{d\Delta\phi_0}{2\pi}\,S(\chi_0,\Delta\phi_0,x),}[EqnPSavg]
appearing in the definition of the surface-subtracted Stokes parameters \eqref{EqnStokesDiff} satisfy the same differential equations \eqref{EqnPSdiff} but with averaged boundary conditions, which are then independent of $\chi_0$.  Hence \eqref{EqnPSavg} are independent of $\chi_0$ and can thus be evaluated at the most convenient value.  With the definitions \eqref{EqnPS}, choosing $\chi_0=0$ for $\bar{P}(\chi_0=0,x)$ and $\chi_0=\pi/2$ for $\bar{S}(\chi_0=\pi/2,x)$ leads to
\eqna{
\bar{P}(x)&=\sin^2\left[\left.\chi(x)\right|_{\chi(1)=0,\Delta\phi(1)=0}\right]=P_{a\to\gamma}(x),\\
\bar{S}(x)&=\sin\left[\left.\chi(x)\right|_{\chi(1)=\pi/2,\Delta\phi(1)=0}\right]e^{-i\!\left.\delta\phi(x)\right|_{\chi(1)=\pi/2,\Delta\phi(1)=0}}-1\\
&=\sqrt{1-P_{a\to\gamma}(x)}e^{-i\delta\phi_{\gamma\to a}(x)}-1,
}[EqnPSavgCP]
which are written explicitly in terms of the ALP-to-photon conversion probability at a distance $xr_0$ for pure ALP initial state
\eqn{P_{a\to\gamma}(x)=\sin^2\left[\left.\chi(x)\right|_{\chi(1)=0,\Delta\phi(1)=0}\right],}[EqnConvProb]
and the photon phase difference at a distance $xr_0$ for a pure photon initial state
\eqn{\delta\phi_{\gamma\to a}(x)=\left.\delta\phi(x)\right|_{\chi(1)=\pi/2,\Delta\phi(1)=0}.}[EqnPhaseDiff]
In \eqref{EqnPSavgCP}, the boundary condition $\Delta\phi(1)=0$ is chosen to avoid singularities.  The photon phase difference \eqref{EqnPhaseDiff} can itself be divided into two independent contributions,
\eqn{\delta\phi_{\gamma\to a}(x)=\delta\phi_\gamma(x)+\delta\phi_a(x),}[EqnPhaseDiffPhotonALP]
with $\delta\phi_\gamma(x)$ the standard birefringence contribution and $\delta\phi_a(x)$ the ALP contribution to the phase difference.  From \eqref{EqnEE} and \eqref{EqnDeltaPhi}, their evolution equations are
\eqn{
\begin{gathered}
\frac{d\delta\phi_\gamma(x)}{dx}=(\Delta_\perp-\Delta_\parallel)r_0,\\
\frac{d\delta\phi_a(x)}{dx}=\Delta_Mr_0\{\cot[2\chi(x)]+\csc[2\chi(x)]\}\sin[\Delta\phi(x)],
\end{gathered}
}[EqnEEPhaseDiff]
with the proper boundary conditions.


\subsection{Stokes Parameters Again}

As a consequence of the above, the surface-subtracted Stokes parameters \eqref{EqnStokesDiff} can be re-expressed as
\eqn{
\begin{gathered}
I(\chi_0,x)-I(\chi_0,1)=A^2\cos(2\chi_0)P_{a\to\gamma}(x),\\
Q(\chi_0,x)-Q(\chi_0,1)=-A^2\cos(2\chi_0)P_{a\to\gamma}(x),\\
U(\chi_0,\delta\phi_0,x)-U(\chi_0,\delta\phi_0,1)=2A_\perp A\sin(\chi_0)\Im\left\{\sqrt{1-P_{a\to\gamma}(x)}e^{-i\delta\phi_0-i\delta\phi_{\gamma\to a}(x)}-e^{-i\delta\phi_0}\right\},\\
V(\chi_0,\delta\phi_0,x)-V(\chi_0,\delta\phi_0,1)=2A_\perp A\sin(\chi_0)\Re\left\{\sqrt{1-P_{a\to\gamma}(x)}e^{-i\delta\phi_0-i\delta\phi_{\gamma\to a}(x)}-e^{-i\delta\phi_0}\right\},
\end{gathered}
}[EqnStokesDiffSoln]
and the Stokes parameters at $x>1$ become
\eqn{
\begin{gathered}
I(x)=I(1)+\left[I_a(1)-\frac{I(1)-Q(1)}{2}\right]P_{a\to\gamma}(x),\\
Q(x)=Q(1)-\left[I_a(1)-\frac{I(1)-Q(1)}{2}\right]P_{a\to\gamma}(x),\\
U(x)=\{U(1)\cos[\delta\phi_{\gamma\to a}(x)]-V(1)\sin[\delta\phi_{\gamma\to a}(x)]\}\sqrt{1-P_{a\to\gamma}(x)},\\
V(x)=\{V(1)\cos[\delta\phi_{\gamma\to a}(x)]+U(1)\sin[\delta\phi_{\gamma\to a}(x)]\}\sqrt{1-P_{a\to\gamma}(x)},
\end{gathered}
}[EqnStokesSoln]
when written in terms of the Stokes parameters and the ALP intensity at the surface.

The fact that the photon phase difference \eqref{EqnPhaseDiffPhotonALP} can be expressed as the expected birefringence contribution---which is independent of the ALP-photon coupling constant---and the ALP contribution, implies that the Stokes parameters at $x>1$ have a simple dependence on their values at $x>1$ without ALPs (the expected astrophysics values, \textit{i.e.}\ the values in the limit of vanishing ALP-photon coupling constant) and the ALP parameters.  Indeed, the photon phase difference is the same than the one due to astrophysics (from birefringence in the magnetosphere) $\delta\phi_\gamma(x)$ plus an ALP contribution $\delta\phi_a(x)$ such that the Stokes parameters \eqref{EqnStokesSoln} at $x>1$ become
\eqn{
\begin{gathered}
I_\text{obs}=I_\text{astro}+\left[I_a(1)-\frac{I_\text{astro}-Q_\text{astro}}{2}\right]P_{a\to\gamma},\\
Q_\text{obs}=Q_\text{astro}-\left[I_a(1)-\frac{I_\text{astro}-Q_\text{astro}}{2}\right]P_{a\to\gamma},\\
U_\text{obs}=[U_\text{astro}\cos(\delta\phi_a)-V_\text{astro}\sin(\delta\phi_a)]\sqrt{1-P_{a\to\gamma}},\\
V_\text{obs}=[V_\text{astro}\cos(\delta\phi_a)+U_\text{astro}\sin(\delta\phi_a)]\sqrt{1-P_{a\to\gamma}}.
\end{gathered}
}[EqnStokesSolnObs]
Equations \eqref{EqnStokesSoln} and \eqref{EqnStokesSolnObs} offer solutions to the Stokes parameters that depend on the surface values or the astrophysical values (again, the values that would be observed in the limit of vanishing ALP-photon coupling constant), respectively, and the ALP parameters through three quantities: the ALP intensity, the ALP-to-photon conversion probability \eqref{EqnConvProb} and the ALP-induced birefringence \eqref{EqnPhaseDiffPhotonALP}.  As expected, the observed quantities in \eqref{EqnStokesSoln} and \eqref{EqnStokesSolnObs} match the surface quantities (modulo the photon birefringence) and the astrophysical quantities, respectively, when ALPs do not mix, \textit{i.e.}\ when both the ALP-to-photon conversion probability and the ALP-induced birefringence vanish.

It is important to point out that the randomness of the phase difference at the surface $\Delta\phi_0$ due to the different origins of the ALP and photon production mechanisms, which allows for averages over $\Delta\phi_0$, ultimately leads to analytic solutions of the polarization quantities at $x>1$ that depend only on the polarization quantities at the surface, $I_a(1)$, $P_{a\to\gamma}(x)$ and $\delta\phi_{\gamma\to a}(x)$.  A common origin for ALP and photon production would not allow for averages over $\Delta\phi_0$, greatly complicating the problem.  Moreover, the separation of the astrophysical and ALP contributions to birefringence allows to disentangle the astrophysical and ALP contributions to the Stokes parameters at $x>1$.  Therefore, a good theoretical understanding of the astrophysics at play could help elucidate the allowed ALP parameter space.


\subsection{Total Intensity and Degrees of Polarization}

Three invariant quantities, \textit{i.e.}\ quantities that are independent of the direction of the coordinate system used to measure them, can be defined from the Stokes parameters.  They are the total intensity $I$, the degree of (total) polarization $p^2$ and the degree of circular polarization $p_c^2$.\footnote{Another invariant quantity, the degree of linear polarization $p_l^2$, is not independent since $p_l^2=p^2-p_c^2$.  Obviously, the non-invariant polarization angles $2\psi=\arctan(U/Q)$ and $2\chi=\arctan(V/\sqrt{Q^2+U^2})$ can also be determined easily from the Stokes parameters.}  In terms of the Stokes parameters, they correspond to
\eqn{I(x),\qquad\qquad p(x)^2=\frac{Q(x)^2+U(x)^2+V(x)^2}{I(x)^2},\qquad\qquad p_c(x)^2=\frac{V(x)^2}{I(x)^2},}[EqnPolInv]
or
\eqn{
\begin{gathered}
I(x)=\left[1-\frac{1}{2}(1-q-2i_a)P_{a\to\gamma}(x)\right]I(1),\\
p(x)^2=\frac{p(1)^2-[p(1)^2-q(1-2i_a)]P_{a\to\gamma}(x)+\frac{1}{4}(1-q-2i_a)^2P_{a\to\gamma}(x)^2}{\left[1-\frac{1}{2}(1-q-2i_a)P_{a\to\gamma}(x)\right]^2},\\
p_c(x)^2=\frac{\left\{\cos[\delta\phi_{\gamma\to a}(x)]+\sqrt{\frac{p(1)^2-q^2}{p_c(1)^2}-1}\sin[\delta\phi_{\gamma\to a}(x)]\right\}^2}{\left[1-\frac{1}{2}(1-q-2i_a)P_{a\to\gamma}(x)\right]^2}[1-P_{a\to\gamma}(x)]p_c(1)^2,
\end{gathered}
}[EqnIntPolDeg]
when expressed in terms of the parameters at the surface \eqref{EqnStokesSoln}, or
\eqn{
\begin{gathered}
I_\text{obs}=\left[1-\frac{1}{2}(1-q-2i_a)P_{a\to\gamma}\right]I_\text{astro},\\
p_\text{obs}^2=\frac{p_\text{astro}^2-[p_\text{astro}^2-q(1-2i_a)]P_{a\to\gamma}+\frac{1}{4}(1-q-2i_a)^2P_{a\to\gamma}^2}{\left[1-\frac{1}{2}(1-q-2i_a)P_{a\to\gamma}\right]^2},\\
p_{c,\text{obs}}^2=\frac{\left[\cos(\delta\phi_a)+\sqrt{\frac{p_\text{astro}^2-q^2}{p_{c,\text{astro}}^2}-1}\sin(\delta\phi_a)\right]^2}{\left[1-\frac{1}{2}(1-q-2i_a)P_{a\to\gamma}\right]^2}(1-P_{a\to\gamma})p_{c,\text{astro}}^2,
\end{gathered}
}[EqnIntPolDegObs]
when expressed in terms of the astrophysical parameters \eqref{EqnStokesSolnObs}.  Here
\eqn{q=\frac{Q(1)}{I(1)}=\frac{Q_\text{astro}}{I_\text{astro}},\qquad\qquad i_a=\frac{I_a(1)}{I(1)}=\frac{I_a(1)}{I_\text{astro}},}[Eqndimless]
when written as functions of the surface and astrophysical quantities, respectively.

Interestingly, the observed invariant quantities \eqref{EqnIntPolDeg} far from the surface ($x>1$) are not solely expressed in terms of the invariant quantities at the surface.  Indeed, for $x>1$ they depend on $Q(1)$ through \eqref{Eqndimless}.  The same is true for the observed invariants \eqref{EqnIntPolDegObs} with respect to their astrophysical values.  This can be understood by the fact that the magnetar dictates a preferred coordinate system with fixed axes for the parallel and perpendicular directions.


\section{Polarization: Perturbative Regime}\label{SecPolPert}

At this point, the effect of ALPs on the polarization parameters \eqref{EqnStokesSoln} or \eqref{EqnIntPolDeg} are completely encoded in the ALP-to-photon conversion probability \eqref{EqnConvProb} and the photon phase difference \eqref{EqnPhaseDiffPhotonALP}.  It is therefore of interest to look for approximate solutions of the ALP-to-photon conversion probability and the photon phase difference when the ALP-photon coupling is small \cite{Raffelt:1987im}.


\subsection{Perturbation Theory}

Without loss of generality, the differential equations \eqref{EqnDiffMat} can be restated as
\eqn{i\frac{d}{dx}A(x)=H(x)A(x)+\delta H(x)A(x),}[EqnDiffMatAH]
where $A=(a\quad E_\parallel\quad E_\perp)^T$ and
\eqn{
H=\left(\begin{array}{ccc}\omega r_0+\Delta_ar_0&0&0\\0&\omega r_0+\Delta_\parallel r_0&0\\0&0&\omega r_0+\Delta_\perp r_0\end{array}\right),\qquad\qquad\delta H=\left(\begin{array}{ccc}0&\Delta_Mr_0&0\\\Delta_Mr_0&0&0\\0&0&0\end{array}\right).
}[EqnAH]
If $\delta H(x)$ is small, it can be considered a perturbation and a standard expansion applied on \eqref{EqnDiffMatAH} leads to the approximate solution $A(x)=\mathcal{U}(x)A(1)$ where the evolution operator is
\eqna{
\mathcal{U}(x)&=\mathcal{U}_0(x)\left[1+(-i)\int_1^xdx'\,\mathcal{U}_0(x')^\dagger\delta H(x')\mathcal{U}_0(x')\right.\\
&\phantom{=}\qquad\left.+(-i)^2\int_1^xdx'\,\mathcal{U}_0(x')^\dagger\delta H(x')\mathcal{U}_0(x')\int_1^{x'}dx''\,\mathcal{U}_0(x'')^\dagger\delta H(x'')\mathcal{U}_0(x'')+\cdots\right],
}[EqnU]
to second order in the perturbation.\footnote{We note that it is necessary to go to second order to verify the probability conservation property to lowest non-trivial order in the ALP-to-photon conversion probability.}  Here, the unperturbed evolution operator is
\eqn{\mathcal{U}_0(x)=\exp\left[-i\int_1^xdx'\,H(x')\right],}[EqnU0]
as expected.

Since the ALP-to-photon conversion probability is defined by
\eqn{P_{a\to\gamma}(x)=|[\mathcal{U}(x)]_{21}|^2,}
the approximate solution to \eqref{EqnConvProb} is
\eqn{P_{a\to\gamma}(x)=\left|\int_1^xdx'\,\Delta_M(x')r_0\,\text{exp}\left\{-i\int_1^{x'}dx''\,[\Delta_a-\Delta_\parallel(x'')]r_0\right\}\right|^2.}[EqnConvProbApprox]
Analogously, the complex exponential of the photon phase difference can be obtained from
\eqn{e^{-i\delta\phi_{\gamma\to a}(x)}=\sqrt{1-P_{a\to\gamma}(x)}\frac{[\mathcal{U}(x)]_{33}}{[\mathcal{U}(x)]_{22}},}
such that the approximate solutions to \eqref{EqnPhaseDiff} and \eqref{EqnPhaseDiffPhotonALP} become
\eqn{
\begin{gathered}
\delta\phi_\gamma(x)=\int_1^xdx'\,[\Delta_\perp(x')-\Delta_\parallel(x')]r_0,\\
\delta\phi_a(x)=-\Im\left\{\int_1^xdx'\,\int_1^{x'}dx''\,\Delta_M(x')r_0\,\Delta_M(x'')r_0\,\text{exp}\left\{-i\int_{x''}^{x'}dx'''\,[\Delta_a-\Delta_\parallel(x''')]r_0\right\}\right\},
\end{gathered}
}[EqnPhaseDiffApprox]
which match the standard birefringence result without ALP perturbation $\delta\phi_\gamma(x)$ [see \eqref{EqnEEPhaseDiff}] plus an ALP contribution $\delta\phi_a(x)$.


\subsection{Analytical Approximations}

In the large conversion radius limit, where the (dimensionless) conversion radius is given by
\eqn{x_{a\to\gamma}=\frac{r_{a\to\gamma}}{r_0}=\left(\frac{7\alpha}{45\pi}\right)^{1/6}\left(\frac{\omega}{m_a}\frac{B_0}{B_c}|\sin\theta|\right)^{1/3},}[EqnConvRadius]
the ALP-to-photon conversion probability \eqref{EqnConvProbApprox} at infinity can be approximated further by
\eqn{P_{a\to\gamma}=(\Delta_{M0}r_0)^2\left|\int_1^\infty dx\,\frac{1}{x^3}\,\text{exp}\left[-i\Delta_ar_0\left(x-\frac{x_{a\to\gamma}^6}{5x^5}\right)\right]\right|^2,}[EqnConvProbApproxB]
since the integral in the exponential is dominated by the region around the (large) conversion radius where $\hat{q}_\parallel\to1$, \textit{i.e.}\ where the magnetic field is dipolar $B(x)=B_0/x^3$.

Although the limit above can be used for the ALP contribution to birefringence, such a simplification cannot be made for the usual birefringence contribution to the photon phase difference since \eqref{EqnPhaseDiffApprox} has non-negligible contributions close to the magnetar surface.  However, in the small surface magnetic field limit we still have that $\hat{q}_\perp\to1$ and $\hat{q}_\parallel\to1$ and the photon phase difference at infinity can be simplified to
\eqn{
\begin{gathered}
\delta\phi_\gamma\approx-\frac{\alpha}{150\pi}\omega r_0\frac{B_0^2}{B_c^2}\sin^2\theta,\\
\delta\phi_a=-(\Delta_{M0}r_0)^2\Im\left\{\int_1^\infty dx\,\int_1^xdx'\,\frac{1}{x^3x'^3}\,\text{exp}\left[-i\Delta_ar_0\left(x-x'-\frac{x_{a\to\gamma}^6}{5x^5}+\frac{x_{a\to\gamma}^6}{5x'^5}\right)\right]\right\},
\end{gathered}
}[EqnPhaseDiffApproxB]
where $\delta\phi_\gamma$---which can be evaluated explicitly---is computed in the small magnetic field limit and $\delta\phi_a$ is computed in the large conversion radius limit.

Although both ALP contributions, \textit{i.e.}\ the ALP-to-photon conversion probability $P_{a\to\gamma}$ \eqref{EqnConvProbApproxB} and the ALP-induced birefringence $\delta\phi_a$ \eqref{EqnPhaseDiffApproxB}, are not analytic yet, they can be further simplified in the large conversion radius limit.  Indeed, they can be rewritten as
\eqn{
\begin{gathered}
P_{a\to\gamma}=\left(\frac{\Delta_{M0}r_0^3}{r_{a\to\gamma}^2}\right)^2\left|\int_\frac{r_0}{r_{a\to\gamma}}^\infty dt\,\frac{1}{t^3}\,\text{exp}\left[-i\Delta_ar_{a\to\gamma}\left(t-\frac{1}{5t^5}\right)\right]\right|^2,\\
\delta\phi_a=-\left(\frac{\Delta_{M0}r_0^3}{r_{a\to\gamma}^2}\right)^2\Im\left\{\int_\frac{r_0}{r_{a\to\gamma}}^\infty dt\,\int_\frac{r_0}{r_{a\to\gamma}}^tdt'\,\frac{1}{t^3t'^3}\,\text{exp}\left[-i\Delta_ar_{a\to\gamma}\left(t-t'-\frac{1}{5t^5}+\frac{1}{5t'^5}\right)\right]\right\},
\end{gathered}
}[EqnALPApprox]
where both integrals may be approximated by integrating from the origin to infinity since they are negligible in the interval $[0,r_0/r_{a\to\gamma}]$ due to the highly oscillatory $1/(5t^5)$ and $1/(5t'^5)$ terms.  Hence in this limit \eqref{EqnALPApprox} can be thought of as functions of one variable, namely $\Delta_ar_{a\to\gamma}$.  By taking $t\to\frac{|\Delta_ar_{a\to\gamma}/5|^\frac{1}{5}}{t}$ and analogously for $t'$, the integrals in the $\Delta_ar_{a\to\gamma}\to0$ limit can be performed analytically, leading to
\eqn{P_{a\to\gamma}\approx\left(\frac{\Delta_{M0}r_0^3}{r_{a\to\gamma}^2}\right)^2\frac{\Gamma\!\left(\frac{2}{5}\right)^2}{25|\Delta_ar_{a\to\gamma}/5|^\frac{4}{5}},\qquad\qquad\delta\phi_a\approx-\left(\frac{\Delta_{M0}r_0^3}{r_{a\to\gamma}^2}\right)^2\frac{\sqrt{5+2\sqrt{5}}\,\Gamma\!\left(\frac{2}{5}\right)^2}{50|\Delta_ar_{a\to\gamma}/5|^\frac{4}{5}}.}[EqnALPApproxAnal]

\begin{figure}[!t]
\centering
\resizebox{15cm}{!}{
\includegraphics{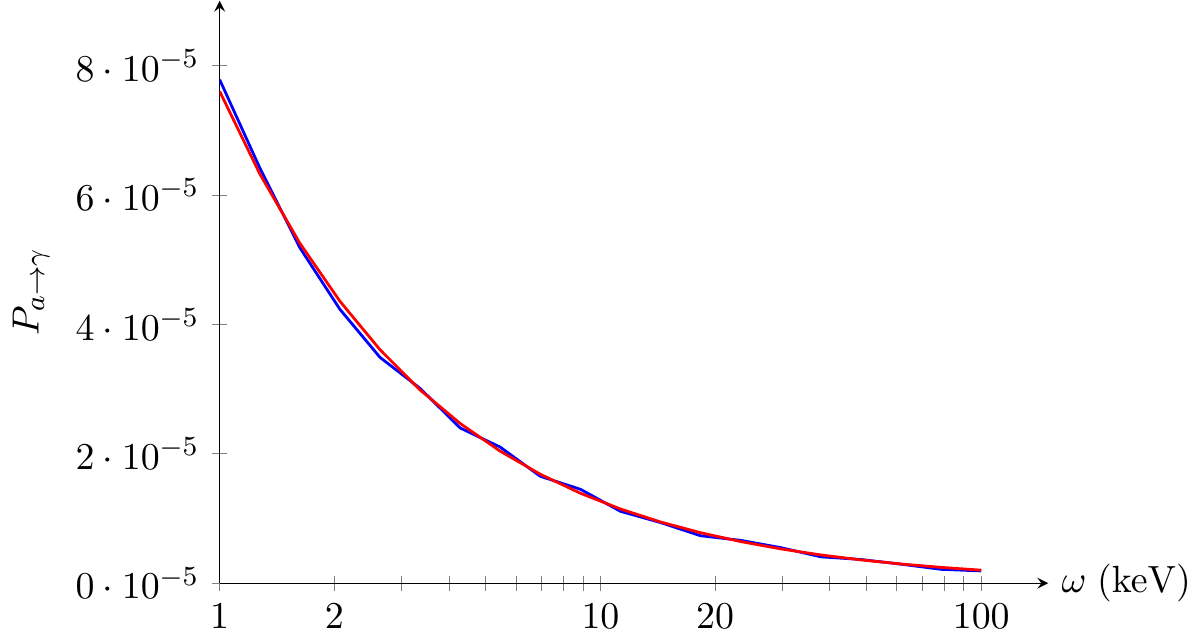}
\hspace{2cm}
\includegraphics{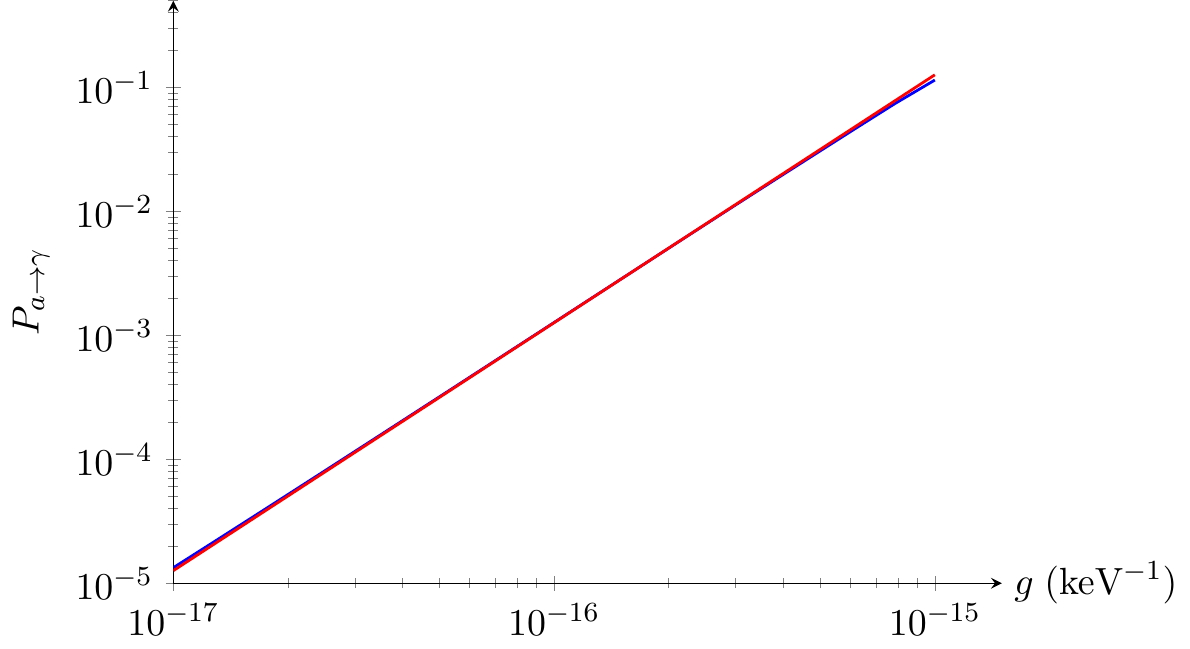}
}
\resizebox{15cm}{!}{
\includegraphics{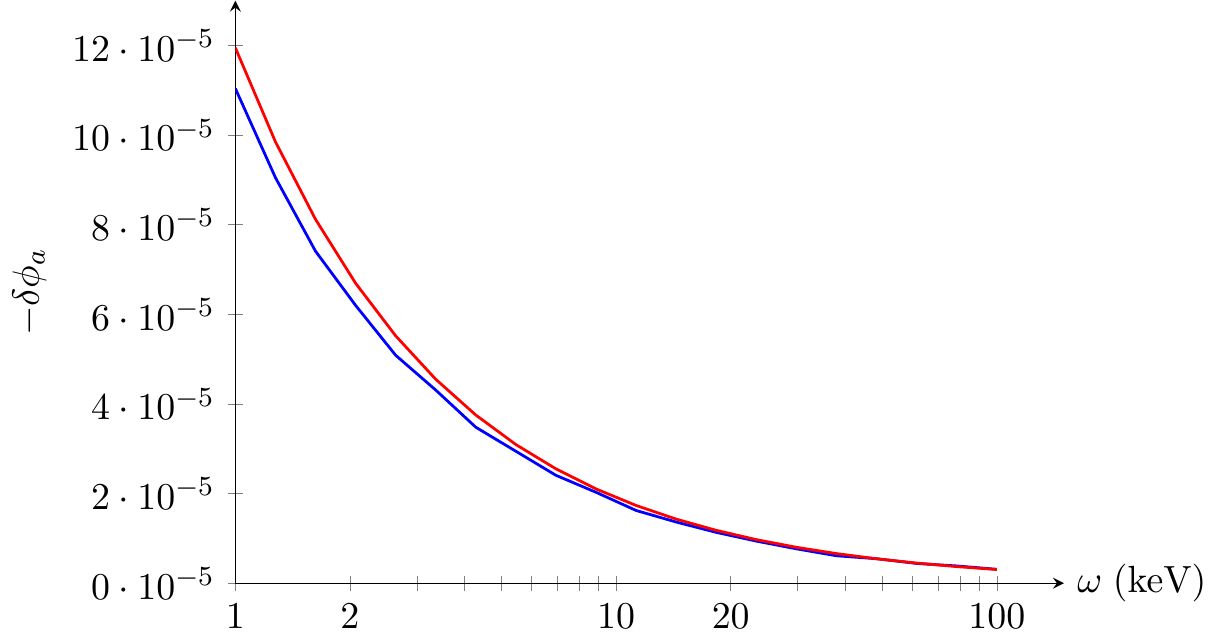}
\hspace{2cm}
\includegraphics{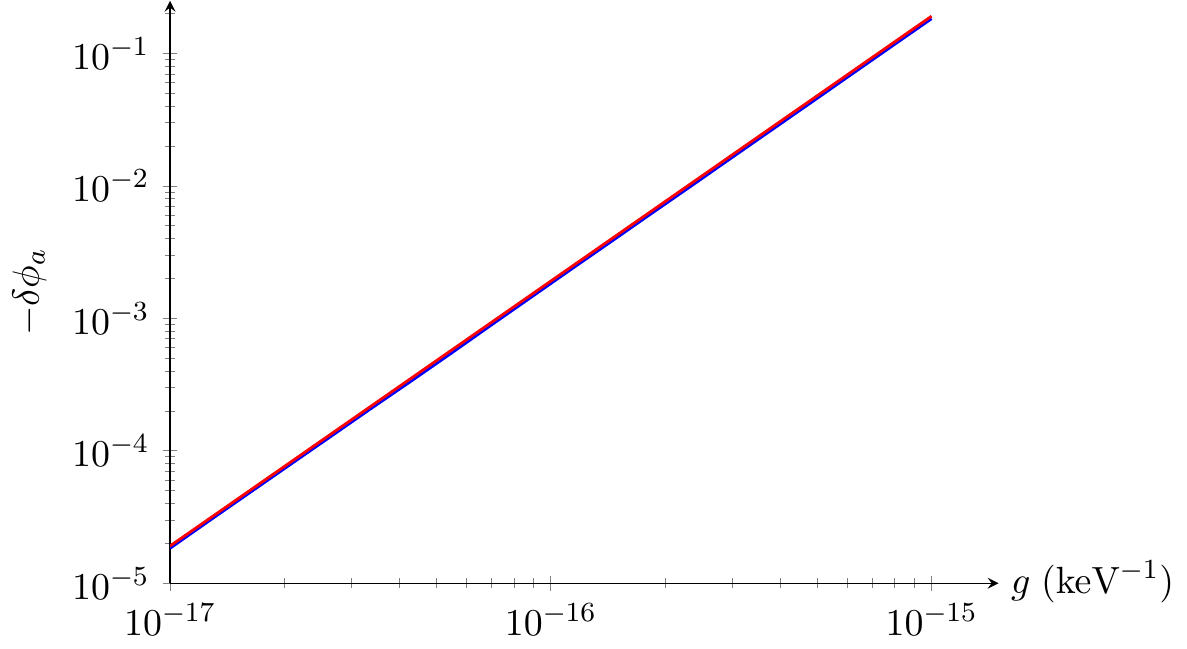}
}
\caption{ALP-to-photon conversion probability and ALP-induced birefringence as a function of $\omega$ and $g$.  The blue lines are derived from the evolution equations \eqref{EqnEE} while the red lines are generated with the help of the approximations \eqref{EqnAllInt}.  The benchmark values of the magnetar parameters are $m_a=10^{-8}\,\text{keV}$, $r_0=10\,\text{km}$, $B_0=10^{12}\,\text{G}$ and $\theta=\pi/2$.  We chose $g=10^{-17}\,\text{keV}^{-1}$ for the left panels and $\omega=10\,\text{keV}$ for the right panels.}
\label{FigALP}
\end{figure}

As a consequence of \eqref{EqnPhaseDiffApproxB} and \eqref{EqnALPApproxAnal}, the relevant quantities can be expressed as
\eqn{
\begin{gathered}
P_{a\to\gamma}=\left(\frac{\Delta_{M0}r_0^3}{r_{a\to\gamma}^2}\right)^2\frac{\Gamma\!\left(\frac{2}{5}\right)^2}{25|\Delta_ar_{a\to\gamma}/5|^\frac{4}{5}}J_P(b_0,|\Delta_ar_{a\to\gamma}/5|^\frac{1}{5},x_{a\to\gamma}),\\
\delta\phi_\gamma=-\frac{\alpha}{150\pi}\omega r_0\frac{B_0^2}{B_c^2}\sin^2\theta\,J_\gamma(b_0),\\
\delta\phi_a=-\left(\frac{\Delta_{M0}r_0^3}{r_{a\to\gamma}^2}\right)^2\frac{\sqrt{5+2\sqrt{5}}\,\Gamma\!\left(\frac{2}{5}\right)^2}{50|\Delta_ar_{a\to\gamma}/5|^\frac{4}{5}}J_a(b_0,|\Delta_ar_{a\to\gamma}/5|^\frac{1}{5},x_{a\to\gamma}),
\end{gathered}
}[EqnAllInt]
where the different $J$-functions are dimensionless integrals given by
\eqn{
\begin{gathered}
J_P(b_0,\xi,\zeta)=\left|\frac{5}{\Gamma\!\left(\frac{2}{5}\right)}\int_0^{\xi\zeta}dx\,x\,\text{exp}\left\{-i\int_1^xdx'\,5x'^4\left[\hat{q}_\parallel\left(\frac{b_0x'^3}{\xi^3\zeta^3}\right)+\frac{\xi^6}{x'^6}\right]\right\}\right|^2,\\
J_\gamma(b_0)=\int_0^1dx\,5x^4\left[\frac{7}{3}\hat{q}_\parallel(b_0x^3)-\frac{4}{3}\hat{q}_\perp(b_0x^3)\right],\\
J_a(b_0,\xi,\zeta)=\frac{50}{\sqrt{5+2\sqrt{5}}\,\Gamma\!\left(\frac{2}{5}\right)^2}\Im\left\{\int_0^{\xi\zeta}dx\,\int_x^{\xi\zeta}dx'\,xx'\,\text{exp}\left\{-i\int_{x'}^xdx''\,5x''^4\left[\hat{q}_\parallel\left(\frac{b_0x''^3}{\xi^3\zeta^3}\right)+\frac{\xi^6}{x''^6}\right]\right\}\right\}.
\end{gathered}
}[EqnJ]
Here it is understood that $\hat{q}_\parallel(b)$ and $\hat{q}_\perp(b)$ are functions of the proper integration variable through their dependence on the magnetic field $b=B/B_c$, which is assumed dipolar.  A comparison between the values obtained from evolving numerically \eqref{EqnEE} and the approximations \eqref{EqnAllInt} for the ALP-to-photon conversion probability and the ALP-induced birefringence is presented in Figure~\ref{FigALP}.

From their definitions, \eqref{EqnJ} tend to one in the appropriate limits, mainly the large conversion radius limit for $J_P$ and $J_a$, and the small surface magnetic field limit for $J_\gamma$.  As such, they can be set to one in \eqref{EqnAllInt} when these limits are verified, leading back to the approximations \eqref{EqnPhaseDiffApproxB} and \eqref{EqnALPApproxAnal}, as expected.

Finally, it is important to point out that it is possible to remain agnostic with respect to the astrophysical origins of polarization when using the Stokes parameters \eqref{EqnStokesSolnObs} and degrees of polarization \eqref{EqnIntPolDegObs}.  Indeed, since they are expressed in terms of the associated astrophysical (non-ALP) quantities, these equations do not depend in any way on $\delta\phi_\gamma$ as given in \eqref{EqnAllInt}.  In other words, ALP contributions are completely disentangled, as mentioned before.  Hence, theoretical astrophysics---from which one can understand the magnetar, its plasma, and the magnetosphere---determines the expected observed polarization quantities after the particles travel from the magnetar surface, through the centimeter-thick plasma where they may undergo mode conversion from resonance \cite{Lai:2006af}, and finally out of the magnetosphere to the observer.  Such quantities are the inputs of \eqref{EqnStokesSolnObs} and \eqref{EqnIntPolDegObs}, but also the outputs of \eqref{EqnStokesSolnObs} and \eqref{EqnIntPolDegObs} when ALP-photon mixing is turned off.  The effects of ALPs are packaged into three important quantities: the ALP intensity, the ALP-to-photon conversion probability, and the ALP-induced birefringence.  The latter two both depend on the ALP and magnetar parameters, mainly the particle's energy $\omega$, the ALP mass $m_a$, the ALP-photon coupling constant $g$, the magnetar's radius $r_0$, the dimensionless surface magnetar magnetic field $b_0=B_0/B_c$, and the angle between the particle's direction of propagation and the magnetic field $\theta$.  Since the presence of ALPs is only significant around the conversion radius \eqref{EqnConvRadius} where ALP-to-photon conversion is non-negligeable and the conversion radius is so large for magnetars (several hundred times the magnetar radius), the dipolar approximation to the magnetic field is warranted and the approximations \eqref{EqnALPApproxAnal} are usually reliable.


\subsection{Stokes Parameters in Perturbative Regime}

The results above are valid in the perturbative regime where the ALP-photon coupling constant is small, which implies $P_{a\to\gamma}\ll1$ and $\delta\phi_a\ll1$.  Therefore the Stokes parameters $U$ and $V$ in \eqref{EqnStokesSolnObs} are barely modified by the presence of ALPs, contrary to the Stokes parameters $I$ and $Q$ that can change wildly due to their dependence on the surface ALP intensity $I_a(1)$, which can be several orders of magnitude larger than the total photon intensity \cite{Fortin:2018aom}.  As a consequence, modifications to the degrees of polarization \eqref{EqnIntPolDegObs} can also be quite substantial when ALPs are considered.

Indeed, since $|q|\leq1$, $0\leq p_\text{astro}^2\leq1$ and $0\leq p_{c,\text{astro}}^2\leq1$ while $i_a\geq0$ without an upper bound, the observed total intensity and degrees of polarization \eqref{EqnIntPolDegObs} can be further approximated by
\eqn{
\begin{gathered}
I_\text{obs}\approx(1+p_a)I_\text{astro},\\
p_\text{obs}^2\approx\frac{p_\text{astro}^2-2p_aq+p_a^2}{(1+p_a)^2},\\
p_{c,\text{obs}}^2\approx\frac{p_{c,\text{astro}}^2}{(1+p_a)^2},
\end{gathered}
}[EqnIntPolDegObsApprox]
where
\eqn{p_a=i_aP_{a\to\gamma}=\frac{I_a(1)P_{a\to\gamma}}{I_\text{astro}},}[Eqnpa]
with $p_a\geq0$ determining the size of the ALP contribution to the polarization quantities.  For completeness, from \eqref{EqnIntPolDegObsApprox} the degree of linear polarization $p_l^2=p^2-p_c^2$ is also given by
\eqn{p_{l,\text{obs}}^2\approx\frac{p_{l,\text{astro}}^2-2p_aq+p_a^2}{(1+p_a)^2}.}[EqnLinPolDegObsApprox]

For a fixed astrophysical scenario in the perturbative regime, we observe from \eqref{EqnIntPolDegObsApprox} and \eqref{EqnLinPolDegObsApprox} that the introduction of ALPs increases the total photon intensity and decreases the degree of circular polarization.  The degree of polarization and the degree of linear polarization on the other hand can increase or decrease depending on the astrophysical Stokes parameter $Q_\text{astro}$ and the ALP contribution, but for very large surface ALP intensities $p_a\gg1$, they both tend to one irrespective of their astrophysical values.  Hence very large ALP surface intensities (when compared to the total photon intensity) lead to vanishing degree of circular polarization while both the degree of polarization and the degree of linear polarization become one, implying linearly polarized light in the parallel direction (the O-mode).  In most astrophysical settings, neutrino energy sink arguments imply that the ALP surface intensity is bounded from above, being no larger than the neutrino intensity and thus putting an upper bound on $I_a(1)\lesssim I_\nu(1)$.\footnote{Since $I_\nu(1)/I_\text{astro}\sim\mathcal{O}(10^4-10^5)$ \cite{Beloborodov:2016mmx}, $I_a(1)$ can be up to four or even five orders of magnitude larger than $I_\text{astro}$.}  Demanding such a constraint on the photon intensity puts an upper bound on $p_a\lesssim1$ \cite{Fortin:2018ehg},\footnote{For a degenerate medium found in magnetars, both the ALP nucleon-nucleon bremsstrahlung emission spectrum and the ALP-to-photon conversion probability peak in the X-ray range, leading to $p_a\lesssim1$ in the X-ray range for ALP-photon coupling constant satisfying the CAST bound \cite{CAST:2017uph}.} with corresponding modifications to all types of degrees of polarization as in \eqref{EqnIntPolDegObsApprox}.


\section{Conclusion}\label{SecConc}

Ongoing and future missions aimed at detecting photon polarization anisotropies in the X-ray band from diverse astrophysical sources will gather invaluable observations that could possibly upend fundamental particle physics.  In particular, X-ray photon polarimetry is poised to make its mark on axion-like-particle (ALP) extensions of the Standard Model, where ALPs mix to parallel photons in background magnetic fields.  As such, magnetars, with their extreme magnetic fields, are promising astrophysical sources to study.

In this paper, we completed the polarization analysis of ALP-photon oscillations that occur in the magnetosphere of magnetars by providing analytical expressions for the four Stokes parameters.  We expressed them in terms of their astrophysical values and three ALP-dependent quantities: the ALP intensity, the ALP-to-photon conversion probability, and the ALP-induced birefringence.  We thus completely disentangled the ALP contributions to the four Stokes parameters from the standard astrophysical contributions.  To achieve such a feat, we relied on the independence of the production mechanisms for photons and ALPs, leveraging the aforementioned independence by averaging over the initial ALP-photon phase difference.  We then expressed the resulting Stokes parameters in terms of two auxiliary functions that we showed were independent of the ALP-photon initial mixture, once averaged.  Consequently, this feature implies that we fully unraveled the contributions of ALPs to the analysis of photon polarization, with the resulting Stokes parameters matching the astrophysical Stokes parameters when the ALP-photon coupling is turned off, as expected.  Hence, we may remain agnostic with respect to the astrophysical origin of photon polarization, considering it as an input to ALP-induced photon polarization.

We also used standard perturbation theory to derive approximate analytical expressions for the ALP-to-photon conversion probability and the ALP-induced birefringence.  We compared them with numerical solutions to the full evolution equations, showing a good match in the appropriate limit.  Effectively, these results lead to fully analytical approximations of the four Stokes parameters that bypass the need to numerically solve the full set of evolution equations, which can be time-consuming.

Finally, we studied the resulting polarization invariants: the total intensity, the degree of total polarization, the degree of circular polarization, and the degree of linear polarization.  As expected, we showed that they correspond to the astrophysical ones when the ALP-photon coupling is turned off.  Surprisingly, we also showed that in the general case with non-vanishing ALP-photon coupling, the polarization invariants---which are functions of the ALP intensity, the ALP-to-photon conversion probability, and the ALP-induced birefringence---are dependent on the astrophysical polarization invariants plus the Stokes parameter $Q_\text{astro}$, which is \textit{not} a polarization invariant.  We argued that this peculiar behavior is related to the existence of a preferred reference frame dictated by the magnetar.  Besides, we used our results to determine the behavior of the polarization invariants in the presence of ALPs, showing that in the limit of large ALP contributions (which can occur when ALP production is large even though the ALP-photon coupling is small), any initial circular polarization of astrophysical origin is wiped out.

Our present polarization results will be used to impose limits on the ALP parameter space with IXPE polarization data from two magnetars in a upcoming article \cite{upcoming}.


\ack{
JFF is supported by NSERC.  KS is supported by the U.~S.~Department of Energy grant DE-SC0009956.  The authors would like to thank Ephraim Gau and Fazlollah Hajkarim.  JFF thanks Daniel C\^ot\'e for useful discussions on Stokes parameters.
}


\bibliography{StokesAll}


\end{document}